\documentclass[aps,showpacs,prl,twocolumn,groupedaddress,amsmath,floatfix]{revtex4}
\usepackage{epsf,bm}
\usepackage[dvips]{graphicx}

\newcommand {\be}{\begin{equation}}
\newcommand {\ee}{\end{equation}}
\newcommand {\bea}{\begin{eqnarray}}
\newcommand {\ba}{\begin{align}}
\newcommand {\eea}{\end{eqnarray}}
\newcommand {\ea}{\end{align}}
\newcommand {\bse}{\begin{subequations}}
\newcommand {\ese}{\end{subequations}}

\begin{document}

\title{A {L}evinson theorem for scattering from a {B}ose-{E}instein condensate}
\author{J. Brand, I. H{\"a}ring, and J.-M. Rost}
\affiliation{Max Planck Institute for the Physics of Complex
Systems, N\"othnitzer Stra{\ss}e 38, 01187 Dresden, Germany\\}

\date{\today}

\begin{abstract}

A relation between the number of bound collective excitations of an
atomic Bose-Einstein condensate and the phase shift of elastically
scattered atoms is derived. Within the Bogoliubov model of a weakly
interacting Bose gas this relation is exact and generalises Levinson's
theorem. Specific features of the Bogoliubov
model such as complex-energy and continuum bound states are
discussed and a numerical example is given.

\end{abstract}


\maketitle

Exact results in physics are rare. 
One such result is
Levinson's theorem \cite{levinson49}, relating the number of bound
states of a given potential to the accumulated scattering phase shift
at threshold.  This theorem 
was first formulated for the single-particle Schr\"odinger equation
and in general does not hold for many-body systems. Instead,
a generalisation for multi-channel scattering exists \cite{newton82},
relating the cumulative sum of the phase shifts
of {\em all open channels} to the number of bound states. Little
can be said about phase shifts in individual channels. 

In this Letter, we consider the scattering of
identical atoms from a spherically symmetric, 
weakly-interacting atomic Bose-Einstein condensate (BEC),
held in a finite, localised trapping
potential.  In this situation, the
phase shift $\delta^l(k)$ of $l$-wave scattering at momentum $k=0$ 
can be related to the number $n_c$ of
bound collective excitations of the condensate by
\be \label{eqn:levBose}
 \delta^l(0) =  \pi(n_c + \delta_{l 0} + \sigma/2) ,
\ee
where $\sigma=1$ for a bound state exactly at the threshold for $s$-wave
scattering and $\sigma=0$ otherwise. 
Equation~(\ref{eqn:levBose}) is
the generalisation of Levinson's theorem to the Bogoliubov equations 
describing the excitations of a weakly interacting BEC and
is the main result of this Letter.

Scattering of cold atoms is a fundamental physical process
relevant to
high-precision atomic spectroscopy and quantum information processing.
As temperatures are lowered, condensation of bosonic atoms is
inevitable and the scattering of single atoms from condensates needs
to be understood.  Scattering experiments involving condensates have
already been demonstrated in the context of four-wave mixing
experiments \cite{Deng1999b,vogels02}. With the
development of atomic lasers \cite{Bloch1999a} as coherent matter wave
sources and with the flexibility introduced by trapping and guiding of
cold atoms with micro-fabricated electrical
circuits \cite{cassettari00,ott01}, precision measurement of scattering
properties becomes feasible. Moreover, interferometric measurements
should allow a direct access of the phase shifts and thereby confirm
Levinson's theorem experimentally in contrast to the case of
conventional atomic scattering experiments where usually intensities are
measured only.  Theoretical attention has been given to
identical particle scattering from BECs at low
energy where transparency effects have been
predicted \cite{Wynveen2000a} and at high energy where density
distributions \cite{idziaszek99} and quantum correlations can be
probed \cite{kuklov99}. Very recently, negative time delays in
one-dimensional scattering from atomic BECs have been
predicted \cite{poulsen03}.  

In the following, we will apply a multi-channel scattering formalism
to the Bogoliubov equations and prove the relation (\ref{eqn:levBose})
by contour integration. We will discuss special situations that can
occur like unstable complex-energy collective modes and continuum
bound states. An instructive numerical example of a realistic
scattering situation is given and the role of the condensate and
channel coupling for the Levinson theorem are discussed.

In the standard Bogoliubov approach \cite{Dalfovo99}, the weakly
interacting Bose gas is described by a condensate with a small amount of
coherent quantum depletion and a gas of non-interacting
quasi-particles describing small thermal or externally induced
collective excitations. The quasi-particles are mixtures of particle and hole
excitations as long as they are located in the condensate
region. Outside they just become free particles and can be identified
with elastically scattered identical atoms. The Bogoliubov picture
relies on the number $N$ of condensate atoms being large; indeed,
the rate of inelastic scattering on a BEC decays
exponentially in the Born
approximation as $N$ grows large \cite{idziaszek99}.

The quasi-particle energies $\epsilon_{\nu l}$ and the particle ($u$) and
hole ($v$) amplitudes are obtained by
solving the Bogoliubov equations
for each partial wave $l$ \cite{Wynveen2000a}:
\begin{subequations} \label{eqn:bog1}
\begin{align}
(T_l + V_{11})u_{\nu l}({r}) -  V_{12} v_{\nu l}({r}) &=
(\epsilon_{\nu l} + \mu) u_{\nu l}({r}) \\
(T_l + V_{22})v_{\nu l}({r}) -  V_{21} u_{\nu l}({r}) &=
(-\epsilon_{\nu l} + \mu) v_{\nu l}({r}) ,
\end{align}
\end{subequations}
where $T_l = -\hbar^2/(2 m) \partial_{rr} + l(l+1)/(2 m r^2)$ is the
kinetic energy with the atomic
mass $m$. The potential is the same in both channels $V_{11} = V_{22} =
V_{\mathrm{trap}}(r)+ 2 g n({r})$ and so  is the off-diagonal coupling
$V_{12} = V_{21} = g n({r})$. The condensate density $n(r)$ and the
chemical potential $\mu$ are determined by solutions of the stationary
radial Gross-Pitaevskii equation:
\be \label{eqn:gpe}
\left\{-\frac{\hbar^2}{2 m}  \partial_{rr} + 
V_{\mathrm{trap}}(r) + g N \frac{|\varphi({r})|^2}{4 \pi r^2}
\right\}\varphi({r}) = \mu \varphi({r})\,,
\ee
where $g\equiv 4\pi\hbar^2 a_s/m$, $a_s$ is the $s$-wave scattering
length, and the condensate order parameter $\varphi({r})$ is assumed
real and normalised to one. Here we
make use of the scattering length approximation for the interatomic
interaction and assume zero temperature, where the density
$n(r) = N \varphi({r})^2/(4 \pi r^2)$. Both these assumptions are taken for
simplicity. Generalisation to finite temperature
Hartree-Fock--Bogoliubov schemes and a more elaborate
treatment of interactions are possible and straightforward as long as
the structure of  Eqs.~(\ref{eqn:bog1}) is preserved.
The trapping potential
$V_{\mathrm{trap}}(r)$ is a finite localised
well that falls off to zero sufficiently quickly for large $r$ 
\footnote{\label{endnt:potcon}We require
$
\int_0^{\infty}|V_{\mathrm{trap}}(r)|r^n\,dr<\infty$ with $n=1,2$
as usual in scattering theory.
}
and traps the condensate, leading to a negative chemical potential
$\mu<0$. 
Formally, the Bogoliubov equations (\ref{eqn:bog1})
can be understood as scattering equations for two coupled channels $u$
and $v$. At $\epsilon_{\nu l} + \mu>0$ there is a scattering continuum 
where the $u$ channel is open and the $v$
channel is closed. At large distance $r$, therefore, the hole
amplitude $v_{\nu l}(r)$
decays exponentially while the asymptotic form
\be
 u_{\nu l} \to \sin[kr + l\pi/2 + \delta^l(k)] \quad \text{as} \quad r\to\infty
\ee
of the particle amplitude defines the phase shift
$\delta^l(k)$ of the scattered atom with wavenumber 
$k = \sqrt{2m/\hbar^2(\epsilon_{\nu l} + \mu)}$.
Note that there is a second scattering continuum for
$-\epsilon_{\nu l} + \mu>0$  where the $v$ channel is open. It
is related to the first continuum by the general symmetry
of the  Bogoliubov equations (\ref{eqn:bog1}) that allows one to construct
new solutions by interchanging $u$ and $v$ and simultaneously changing
the sign of $\epsilon$. This property can be traced to the
invariance of the coupling matrix $V$ under  exchange of diagonally
opposite matrix elements
\be \label{eqn:symm}
V_{12}\leftrightarrow V_{21}\quad\text{and}\quad V_{11}
\leftrightarrow V_{22}. 
\ee

We now proceed with the proof of Eq.~(\ref{eqn:levBose}) for the
special case $l=0$. To this end, we slightly generalise the Bogoliubov
equations~(\ref{eqn:bog1}) by introducing the channel momenta $k_1$
and $k_2$ as independent variables and write in matrix notation
\be \label{eqn:2cscat}
(-\partial_{rr}\openone + {V}){\phi} = {K}^2 {\phi},
\ee
where ${K} = \text{diag}(k_1,k_2)$ and  $\phi=[u_\nu(r), v_\nu(r)]^T$. We
have now chosen units where $\hbar^2/(2 m) = 1$ and dropped the channel
index $l$. 
Analytical properties of multi-channel scattering solutions are most
conveniently discussed by analysing the Fredholm determinant
$\Delta(k_1,k_2)$,  which 
generalises the Jost function of single-channel scattering \cite{newton82}.
We define the Fredholm determinant by
$\Delta(k_1,k_2) = \det \cal{F}$ with
the Jost matrix $\cal{F}$
\bea \label{eqn:Jost}
{\cal F} = \openone + \int_0^{\infty} e^{i K r} V(r) \Phi(K,r)\,dr .
\eea
Here, the columns of the matrix $\Phi(K,r)$ are regular solutions
of Eq.~(\ref{eqn:2cscat}).  $\Phi(K,r)$ obeys the 
system of coupled integral equations
\begin{multline}
K \Phi(K,r) =
 \sin(K r)
 + \int_0^r \sin[K(r-r')] V(r')  \Phi(K,r')\,dr' .
\label{eqn:regular}
\end{multline}
The Fredholm determinant
$\Delta(k_1,k_2)$ is an entire function of
$k_1$ and $k_2$ for finite range couplings. Under weaker conditions
\cite{endnote19} 
the function $\Delta(k_1,k_2)$ is still analytic
for $\Im k_1 > 0$ and
$\Im {k_2}> 0$.

The function $\Delta(k_1,k_2)$ contains the complete information about the
physical scattering process and allows one to construct the {\bf
S}~matrix. In particular, if  $k_1$ or $k_2$ are real 
\be\label{eqn:Dphase}
 \Delta(k_1,k_2) = |\Delta| e^{-i\delta}
\ee
is directly related to  the
phase shift $\delta$ of the open channel. Furthermore, we have the
symmetry
\be\label{eqn:Ksymmetry}
\Delta(k_1,k_2) = \Delta^*(-k_1^*,-k_2^*),
\ee
which follows for a real symmetric coupling matrix $V_{ij}$,
and for $\Im k_1 \ge 0$ and  $\Im {k_2} \ge 0$, the relations
\be\label{eqn:Klarge}
\Delta(k_1,k_2)\to 1\quad\text{as}\quad |k_1|,|k_2|\to\infty
\ee
hold \cite{newton82}.
{
The symmetry
\be\label{eqn:symDelta}
\Delta(k_1,k_2) =
\Delta(k_2,k_1) 
\ee
is a special property of the Bogoliubov equations and will be
instrumental for the proof of Eq.~(\ref{eqn:levBose}). Equation
(\ref{eqn:symDelta}) can be derived from the  symmetry
(\ref{eqn:symm}) using Eqs.~(\ref{eqn:Jost}) and (\ref{eqn:regular}).
}

\begin{figure}[ht]
\begin{center}
\epsfxsize=8cm \leavevmode \epsfbox{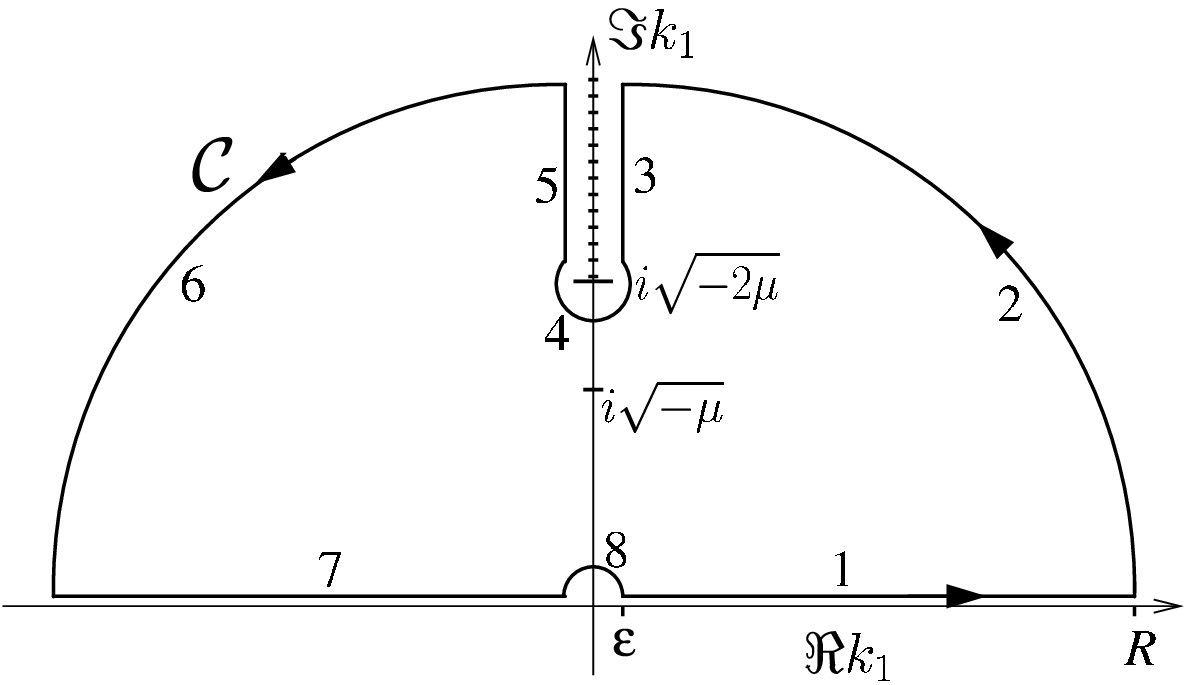}
\caption{\label{fig:1} Analytic structure of $\Delta(k_1)$ in the
upper half of the physical sheet and integration contour ${\cal
C} =\sum_{i=1}^8{\cal C}_i$ as detailed in the text. The segments
${\cal C}_7$, ${\cal C}_1$ 
and  ${\cal C}_3$, ${\cal C}_5$ lie exactly on the real and imaginary
axis, respectively.
}
\end{center}
\end{figure}

We now return to the Bogoliubov equations
(\ref{eqn:bog1}) where the 
channel momenta $k_1$ and $k_2$ are not independent variables but
related through $k_1^2 + k_2^2 = 2 \mu$. Resolving for $k_1$, we
specifically choose
\be \label{eqn:root}
  k_2(k_1) = i \sqrt{k_1^2 - 2\mu},
\ee
where $\sqrt{\cdot}$ denotes the usual positive square root with a
branch cut along the negative real axis. If $k_2$ is given by
(\ref{eqn:root}) we say that $k_1$ is in the physical sheet. The
choice (\ref{eqn:root}) 
assures that $\Im{k_2}>0$, whenever $\Im{k_1}>0$. We see that
$\Delta(k_1) \equiv \Delta(k_1,k_2(k_1))$ is analytic in the upper
complex half plane, except for a branch point at $k_1 = i\sqrt{-
2\mu}$ and a cut along the imaginary axis where $k_1/i\ge
\sqrt{-2 \mu}$ originating from the square root (\ref{eqn:root}) as
shown in Fig.~\ref{fig:1}.
This is exactly where $k_2^2$ becomes positive, i.e.~the $v$
channel is open and supports scattering solutions.

%
%

Bound solutions of the Bogoliubov equations correspond to zeros of
$\Delta(k_1)$ in the upper half of the physical sheet where $\Im k_1\ge 0$.
Solutions with positive
quasiparticle energies $\epsilon_{\nu}$, the usual case for stable
ground state condensates, are found on the imaginary axis at $k_1/i
\in [0, \sqrt{-\mu}]$.
Due to the symmetry (\ref{eqn:symm}) of the
Bogoliubov equations, every bound state at $k_1/i =
\sqrt{-\epsilon_{\nu}-\mu}<\sqrt{-\mu}$ has an image at $k_1/i =
\sqrt{+\epsilon_{\nu}-\mu}>\sqrt{-\mu}$.
Physically,
only the solution with positive quasiparticle energy is meaningful.
At $k_1 = i \sqrt{-\mu}$ corresponding to zero quasiparticle energy
$\epsilon_{0}=0$, a trivial bound solution of the Bogoliubov equations
causes a double zero of $\Delta(k_1)$. This solution with $u_{0} = v_0
\propto \varphi(r)$ is proportional to the condensate order parameter
and does not describe a condensate excitation. In fact, quasiparticle
excitations are confined to the orthogonal complement of
$\varphi(r)$.

following integral in the complex $k_1$-plane over the closed contour ${\cal
C} = \sum_{i=1}^8{\cal C}_i$ shown in Fig.~\ref{fig:1}:
\be \label{eqn:levint}
  \int_{{\cal C}} {d \ln\Delta(k_1)} =  2\pi i n_z\equiv 2\pi i(2 n_c + 2)\,,
\ee
which counts the number $n_z$ of zeros of $\Delta(k_1)$ with their
multiplicity within the region
enclosed by ${\cal C}$.
From the preceeding discussion it is clear, that $n_z$ is given by
twice the number of bound collective 
excitations $n_c$ enclosed by ${\cal C}$ plus the contribution from
the trivial solution.
Note that the contour ${\cal C}$ is more complicated than
in the proof of the ordinary Levinson theorem, due to the cut and the
branch point 
which has to be avoided.

We will now discuss contributions to the integral on the left
hand side of Eq.~(\ref{eqn:levint}). As in the usual proof of the
Levinson theorem, the contributions on the negative
and positive imaginary axis can be related to the scattering phase
shift $\delta$ and with Eqs.~(\ref{eqn:Dphase}) and
(\ref{eqn:Ksymmetry}) we obtain
\be
\int_{{{\cal C}_7} + {\cal C}_1}
d\ln\Delta(k_1)  = 2 i [\delta(\varepsilon) -\delta(R)] .
\ee
The small semicircle ${{\cal C}_8}$ of radius $\varepsilon$ around
the origin gives a
vanishing contribution for small $\varepsilon$ as do the quarter circles
${{\cal C}_2}$  and  ${{\cal C}_6}$ for large $R$ due to
Eq.~(\ref{eqn:Klarge}). 

The remaining contributions of the integrals over ${{\cal C}_3}$,
${{\cal C}_4}$, and ${{\cal C}_5}$ along the imaginary axis can be
related to the contributions along the real axis
due to the special structure of the Bogoliubov equations.
In fact, Eq.~(\ref{eqn:root}) defines a
conformal mapping, which maps the segments ${{\cal C}_3}$, ${{\cal
C}_4}$, and ${{\cal C}_5}$ of the $k_1$ plane onto the segments
${{\cal C}_7}$, ${{\cal C}_8}$, and ${{\cal C}_1}$ of the $k_2$ plane,
respectively. With the symmetry (\ref{eqn:symDelta}) we obtain
$\int_{{\cal C}_3} d\ln\Delta(k_1) = \int_{{\cal C}_7} d\ln\Delta(k_1)$
and likewise $\int_{{\cal C}_4} = \int_{{\cal C}_8}$ and  $\int_{{\cal
C}_5} = \int_{{\cal C}_1}$.

We have now evaluated the left hand side of Eq.~(\ref{eqn:levint}),
which becomes 
$4 i [\delta(0) -\delta(\infty)]$
in the limits $\varepsilon \to 0^+$ and $R\to\infty$. Note that we can
set $\delta(\infty) = 0$ due to Eq.~(\ref{eqn:Klarge}).
Thus we obtain Eq.~(\ref{eqn:levBose}) for
$l=0$ and under the assumption that there are no zeros of
$\Delta(k_1)$ on the real axis. The generalisation of this proof for
$l>0$ is straightforward within the  partial-wave
formalism. However, for $l>0$ {\em all} zeros enclosed by the contour ${\cal
C}$ correspond to collective excitations since the modes $u$ and $v$
are always orthogonal to the condensate wavefunction $\varphi$.

Finally, we consider the possibility of zeros of $\Delta(k_1)$ on the
real axis. Additionally to zeros at the threshold
$k_1 = 0$, also encountered in single-channel scattering, we cannot
exclude the possibility of  zeros of $\Delta(k_1)$ at real $k_1 \ne 0$.
If zeros of $\Delta(k_1)$ on
the real axes exist, they fully contribute to a continuously measured
phase shift in the same way as continuum bound states do in
conventional  multi-channel scattering theory with the exception of a
zero at $k_1=0$ in the $l=0$ channel \cite{newton82}. This specific
case is known as a
half bound state as it only contributes $\pi/2$ to the phase shift,
hence the term $\sigma/2$ in Eq.~(\ref{eqn:levBose}).

So far we have assumed that the condensate
is in a stable ground state of the trap, in which case all bound
excitations described by the Bogoliubov equations have finite pseudo
norm $\eta \equiv \int \left(|u|^2-|v|^2\right) dr \ne 0$ and those
with $\eta >0$ have positive quasi-particle energy
$\varepsilon_{\nu}$. However, this restriction is not necessary. For
excited states $\varphi(r)$ of the stationary Gross-Pitaevskii equation
(\ref{eqn:gpe}), anomalous solutions of the Bogoliubov equation are
possible with $\epsilon_{\nu} \le 0$ and $\eta >0$. Such anomalous
modes are known to occur for solitons in highly elongated traps
\cite{Muryshev1999a,dziarmaga02} and for
vortices \cite{fetter01}. Furthermore, bound solutions with complex
$\epsilon_{\nu}$ and $\eta =0$ may occur, which correspond to zeros of
$\Delta(k_1)$ on the physical sheet but {\em off} the imaginary
axis. These solutions arise when the condensate is in a stationary but
unstable excited state of the Gross-Pitaevskii equation and describe
modes of exponential decay, predicted and seen, e.g., for dark solitons
\cite{Anderson2001a,brand02a} and attractive condensates
\cite{strecker02}.  These unstable complex modes are not to be
confused with scattering resonances, which have zeros below the real
axis and in the unphysical sheet. Both anomalous and complex
Bogoliubov modes contribute to the contour integral (\ref{eqn:levint})
and thus are predicted to be visible in the build up of the elastic
scattering phase shift $\delta(0)$, which can, in principle, be
probed experimentally.

%
\begin{figure}[ht]
\begin{center}
\includegraphics[height=8cm,angle=270]{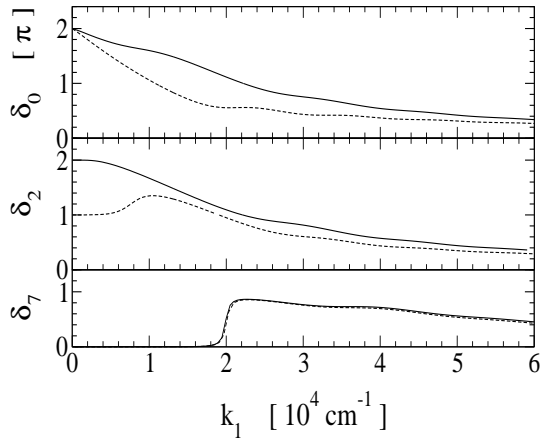}
\caption{\label{fig:2} 
Phase shifts $\delta_l$ from the full solution of Eqs.~(\ref{eqn:bog1})
({\it solid}), simple potential scattering without coupling [$V_{12}=0$ in Eqs.~(\ref{eqn:bog1})] 
({\it dashed}). In the example,
2000 Rubidium atoms are held in a finite harmonic trap of 
frequency $\omega =2 \pi 200 s^{-1}$ and
depth $V_0=8.0 \hbar \omega$  resulting in the chemical potential
$\mu = -3.34\hbar \omega$ and allowing for up to two 
  bound collective excitations for $1\le l\le 5$.
}
\end{center}
\end{figure}
%

This is exemplified in Fig.~\ref{fig:2} for a ground-state BEC of 2000
Rubidium atoms. 
Note that the relevant range of scattering energies corresponding to
temperatures of the order of 10~nK is within the accessible range of
current cold-atom experiments.
For $l=0$ one bound
collective excitation and  the trivial solution account for a phase shift
at threshold $\delta_0(0)$ of $2\pi$ both in the full Bogoliubov equations
(solid) and in the potential
scattering approximation (dashed) where we have set $V_{12}=0$.
Both solutions coincide in the absence of a bound mode for
$\delta_7(k_1)$ which exhibits, instead, a resonance caused by a zero of
$\Delta(k_1)$ with $k_1$ having a small negative imaginary part. The
interesting case is $\delta_2(k_1)$ which indicates one more bound 
state in the 
full solution compared to simple potential scattering.

\begin{figure}[htb]
\includegraphics[height=8cm,angle=270]{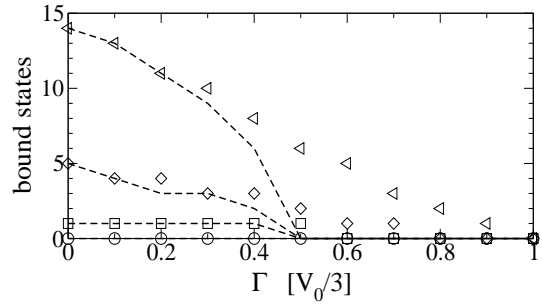}
\caption{\label{fig:3} Number of bound collective modes for a
square-well of depth $V_0 = 10$, 100, $1000\; \hbar^2/ m a^2$,
respectively, and radius $a$ 
with a Thomas-Fermi approximation for the wave function as a function
of the nonlinear coupling $\Gamma = N a_s/a$. Full solution
({\it symbols}), potential scattering ({\it dashed}) as in Fig.~\ref{fig:2}. 
}
\end{figure}

The offdiagonal coupling $V_{12}$ is of attractive nature
and binds extra states 
although the interparticle interaction is repulsive in both examples.
For large scattering momenta and large angular momenta this coupling
to the $v$ mode becomes less important, as the details of the interaction
region are hardly probed. 
Figure~\ref{fig:3} shows the number of bound states of a square well
model system. As the well is filled up with repulsive atoms, fewer
collective excitations are bound. Note that for $\Gamma > V_0/6$ the
trapped BEC has bound quasiparticle states only due to the coupling between
hole and particle modes.

We have presented a Levinson theorem relating the number of bound
collectively excited states of an interacting many-body system, a
BEC, to the phase shifts of single-particle
scattering. With the possibilities of cold atom
scattering and interference, we can expect to see a direct experimental
verification of a fundamental theorem of mathematical
physics. However, our derivation is  based on a weakly
interacting Bose gas without inelastic scattering processes.
It should be seen as a challenge both to experiments
and theory to find the corrections to Eq.~(\ref{eqn:levBose}) in a
real interacting system. Dilute-gas atomic BECs
are an ideal system for this because the experimental set up is
stupendously manageable.

%
%

\bibliographystyle{prsty}

\end{document}